\documentclass[aps,prl,reprint,superscriptaddress,floatfix]{revtex4-1}

\usepackage{slashed}
\usepackage{amsmath}
\usepackage{graphicx,bm}
\usepackage{color}
\usepackage{hyperref}
\usepackage[squaren]{SIunits}

\usepackage[dvipsnames]{xcolor}
\definecolor{darkgreen}{rgb}{0.0, 0.2, 0.13}

\usepackage{verbatim}

\usepackage{grffile} 

\usepackage{caption}
\usepackage{subcaption}

\graphicspath{{./images/}}

\let\vec\mathbf

\begin{document}

\title{Narrow bandwidth gamma comb from nonlinear Compton scattering using the polarization gating technique} 

\author{M. A. Valialshchikov}
\email[]{maksim.valialshchikov@skoltech.ru}
\affiliation{High Performance Computing and Big Data Laboratory, Skolkovo Institute of Science and Technology, Moscow, Russia}

\author{V.~Yu. Kharin}
\noaffiliation

\author{S.~G. Rykovanov}
\email[]{s.rykovanov@skoltech.ru}
\affiliation{High Performance Computing and Big Data Laboratory, 
Skolkovo Institute of Science and Technology, Moscow, Russia}

\date{\today}

\begin{abstract}

Nonlinear Compton scattering is a promising source of bright gamma-rays. Using readily available intense laser pulses to scatter off the energetic electrons, on the one hand, allows to significantly increase the total photon yield, but on the other hand, leads to a dramatic spectral broadening of fundamental emission line as well as its harmonics due to the laser pulse shape induced ponderomotive effects. In this paper we propose to avoid ponderomotive broadening in harmonics by using the polarization gating technique - a well-known method to construct a laser pulse with temporally varying polarization. We show that by restricting harmonic emission only to the region near the peak of the pulse, where the polarization is linear, it is possible to generate a bright narrow bandwidth comb in the gamma region. 

\end{abstract}


\maketitle




Recently, there has been a revival of interest in the Compton photon sources based on scattering of intense laser pulses from relativistic electron beams, which is apparently due to the present day availability and maturing of both compact powerful laser systems and compact laser-plasma based accelerators (LPAs)~\cite{faure2004laser, blumenfeld2007energy, nedorezov2004UFN, geddes2015compact, rykovanov2014quasi}. The main advantage of Compton based photon sources over, for example, bremsstrahlung sources is their monochromaticity allowing their usage in nuclear spectroscopy~\cite{geddes2015compact, albert2011design, bertozzi2008nuclear, quiter2011transmission}, medicine~\cite{carroll2003pulsed, weeks1997compton}, and other applications~\cite{quiter2008method, carpinelli2009compton}. The price one has to pay for this quality is a very low cross-section of the process leading to a quite meager photon brightness. One possible and seemingly straightforward way to increase the source brightness is to increase the intensity of the laser pulses used for scattering, and indeed this leads to a significant enhancement of the total photon yield. However, due to the temporal shape of the laser pulses, ponderomotive effects (the ``slow-down'' of electrons due to the $\mathbf{v}\times\mathbf{B}$ force) start playing an important role in electron dynamics and lead to the so-called ponderomotive spectral broadening~\cite{hartemann1996spectral, hartemann2013nonlinear, rykovanov2016controlling, heinzl2010beam, seipt2011nonlinear}, destroying the main quality of the Compton sources - their monochromaticity.
One may use laser pulses with flat-top profiles to avoid ponderomotive broadening \cite{hartemann1996spectral}, but experimentally it is exceptionally challenging to create such pulses with high intensity. Recently, it was proposed to perfectly compensate the ponderomotive broadening by using properly chirped laser pulses, where pulse frequency is a nonlinear function of time exactly following the change of the laser pulse envelope~\cite{ghebregziabher2013spectral, terzic2014narrow, rykovanov2016controlling, seipt2015narrowband}. It has also been shown theoretically that harmonics of the fundamental Compton line are also narrow when using properly chirped laser pulses~\cite{terzic2016combining}. However, to the best of our knowledge, generating such laser pulses with a nonlinear temporal chirp in the laboratory is extremely challenging. Recently, two papers were published showing that it is theoretically possible to generate a narrow bandwidth spectrum for high intensities using only linear chirp~\cite{kharin2018higher, seipt2019optimizing}, requiring, however, accurate tuning of the experimental setup.

In this paper, we present a simple method to avoid ponderomotive broadening in the harmonics of the fundamental Compton scattering line. For intense laser pulses, harmonics overlap into complete disarray, while with our approach harmonics spectrum forms a well-defined comb. Our idea relies on a method that is very well known in the attosecond community -- to use laser pulses with temporally varying polarization (with circular polarization in the wings and linear polarization only in the middle of the pulse) to gate emission of harmonics only to the part of the pulse where the polarization is linear \cite{rykovanov2008intense}. In this way, one can generate single attosecond pulses instead of a train of attosecond pulses. Just like in the case of gas or surface harmonics, it is well known that intense circularly polarized light does not generate on-axis harmonics in the Compton backscattering from energetic electrons, whereas linearly polarized light produces harmonics of the fundamental Compton line. 


In this paper, we show using theoretical methods and numerical calculations that the polarization gating technique allows one to limit the emission of Compton harmonics only to the peak of the laser pulse where the polarization is close to linear and ponderomotive effects due to the gradient of intensity are lower. And although the main emission line (fundamental Compton harmonic) still suffers from ponderomotive broadening~\cite{hartemann1996spectral, brau2004oscillations, kharin2016temporal}, we show that its harmonics are narrow and bright, hence exhibiting a comb in the gamma-ray region. Throughout the paper, we use units with $\hbar=c=1$, dimensionless spacetime $\left(x\omega_L\rightarrow x\right)$ and energy ($\omega/\omega_L\rightarrow \omega$) variables by rescaling with the central laser frequency $\omega_L$. Dimensionless laser pulse amplitude is given by $a_0=e A/ m$, where $e,m$ are the absolute value of electron charge and electron mass respectively. All numerical simulations were conducted within the classical description of Compton scattering, which is valid when electron recoil and radiation friction could be neglected. Therefore, the recoil parameter should satisfy $\zeta = 2\gamma E_p a_0/ m \ll 1$, where $\gamma$ is the relativistic factor of  the electron, $E_p$ is the energy of the incoming photon in the laboratory frame, and radiation friction could be neglected for $a_0 < \epsilon_{\text{rad}}^{-1/3}$, $\epsilon_{\text{rad}}=2\gamma \frac{4\pi}{3} \frac{r_e}{\lambda_L}$, where $r_e$ is the classical electron radius, $\lambda_L$ is the laser pulse wavelength \cite{nikishov1964quantum, mourou2006optics}. Usually practical applications require $a_0 \sim 1$, and in this range of parameters, all constraints are satisfied \cite{rykovanov2014quasi}. For the chosen parameters, radiation reaction leads to the harmonics redshift and minor broadening (see Supplementary Materials). However, this is beyond the scope of the current work, a detailed investigation of the radiation reaction's impact on electron's motion and radiation distributions could be found in \cite{di2008exact, thomas2012strong, ruijter2018analytical}.


Let us start with a brief description of the polarization gating technique. There are several experimental ways to realize a laser pulse with time-varying ellipticity, all based on linear optics. From the mathematical point of view, polarization gated pulse (PGP) is an overlap of two circularly polarized laser pulses with opposite handedness. One can write the following expression for the vector potential of the PGP, neglecting the carrier-envelope phase effects 

\begin{equation}
    \small
    \vec{A}_\perp(\phi)=\frac{a_0}{2}e^{i\phi}\left( g\left(\phi-\frac{\delta}{2} \right)\boldsymbol{\varepsilon}_+ + 
    g\left(\phi+\frac{\delta}{2} \right)\boldsymbol{\varepsilon}_-
    \right)+c.c.\mathrm{,}
\end{equation}
where the vector potential is made dimensionless with the help of rescaling $e\vec{A}/m \rightarrow \vec{A}$, $\phi=t-z$ is the light-front time, $\delta$ is the normalized delay between two pulses, $\boldsymbol\varepsilon_{\pm}=\left[1, \pm i\right]^T$ is the ellipticity parameter defining left or right handed circular polarization. If we take into account the carrier-envelope phase effects then in order to have linear polarization at $\phi=0$ it is necessary that $\delta=n\pi$ with $n$ an integer number (see Supplementary Materials).

To study the nonlinear Compton scattering, it is convenient to work in the electron frame of reference, where the electron is initially at rest, $p=(m,0,0,0)$. Scattered photon spectrum in the laboratory frame, where the electron is initially counter-propagating the laser pulse with the energy $\gamma m$, can be obtained in a straightforward manner using the Lorentz transformation.

Knowing the laser pulse amplitude from Eq. (1), one may obtain the harmonics on-axis central frequency 

\begin{equation}
    \omega_n(\phi) = \frac{n}{1+A_{\perp}^2(\phi)}\mathrm{,}
\end{equation}
where $n$ is an odd integer and stands for the harmonic number.




In the frame of reference where the electron was initially at rest, the solution of electron's equations of motion in the plane wave field is widely known (in our problem setting, the wave is coming from the $-z$ direction) \cite{esarey1993nonlinear}:

\begin{align}
    \gamma - u_z = 1 \leftrightarrow u_z &= \frac{A_\perp^2}{2}\mathrm{,}\\
    u_\perp &= A_\perp \mathrm{,}
\end{align}
where $u$ is the electron four-velocity.

The distribution of radiation emitted by the electron is given by \cite{jackson1999classical} 
\begin{equation}
    \frac{d^2 I}{d\omega d\Omega} = \frac{\omega^2}{4\pi^2}\left| \int_{-\infty}^{\infty} d\phi \:\: \vec{n} \times [\vec{n} \times \vec{u}] \: e^{i \omega (\phi + z - \vec{n}\vec{r})} \right|^2 \mathrm{,}
    \label{spec_integral}
\end{equation}
where $\vec{n}$ is the direction of observation.

In high intensity fields ($a_0\sim1$) the longitudinal electron motion is strongly modulated by the magnetic field due to the laser pulse temporal envelope. It complicates the analytical description of the process, namely, the calculation of spectrum integral in Eq. (\ref{spec_integral}) could be done only for specific configurations (on-axis spectrum, linear or circular polarization) and leads to the spectral ponderomotive broadening. 
Therefore, to calculate the spectrum for arbitrary ellipticity of the incident pulse or evaluate the spectrum's angular distribution, one needs to calculate this integral numerically. The efficient numerical procedure for the spectrum calculation is widely known \cite{kharin2016temporal}: 1) integrate Eqs. (3)-(4) to obtain the trajectories, 2) proceed from even grid $\{\phi_i\}$ to even grid in retarded time, 3) consider the integral in Eq. (\ref{spec_integral}) as the Fourier transform in retarded time and use Fast Fourier Transform to get the result. 





From now on, we consider a Gaussian temporal envelope with mean $\phi_0$ and length $\tau$: $\exp{[-(\phi-\phi_0)^2/\tau^2]}$. 
Eq. (2) illustrates that for a linearly polarized pulse the emitted frequency at the top of the pulse is the lowest and gradually increases when going to the wings.
By varying the delay ($\delta = 2\phi_0$) between two circular pulses (centered at $-\phi_0$ and $\phi_0$) with opposite handedness with respect to $\tau$ it is possible to control how sharp and bright the emitted harmonics are. 

To investigate the influence of delay variation between two circular pulses, we modeled several PGPs along with their backscattered spectra for different delay parameters. We found out that for the optimal delay $\delta=\tau$, one may observe a narrow and bright gamma comb, while for other delay parameters, the harmonics are either too scarce or overlap into complete disarray (one can find the analytical derivation in Supplementary materials). Figure \ref{fig:A_and_back_spect_for_opt_delay} illustrates the vector potential of a PGP with varying ellipticity and backscattered spectra for the optimal delay $\delta=\tau$. Other laser pulse parameters were as follows: $a_0=3, \tau=8\pi$. To obtain spectrum in ergs, one needs to multiply the values on the figure by the normalization coefficient $e^2 \omega_L$. One can see that the polarization is linear (ellipticity $=0$) at the center of the pulse and circular (ellipticity $=1$) in the wings. The dependence of harmonics generation efficiency (normalized harmonics amplitude) from polarization is ``gaussian-like'': for linear it equals one and then smoothly goes to zero for circular polarization, the sharpness of this transition depends on $a_0$ and harmonics number $n$ - for large laser pulse amplitudes and high harmonics it is much sharper (e.g. for $a_0=1, n=7$ for a rectangular pulse the efficiency of harmonics generation for ellipticity $=0.5$ is only $\sim 0.1$). Therefore, the delay between pulses controls the sharpness of transition from one polarization to another as well. Moreover, the resulting backscattered spectrum shows that it is possible to avoid the ponderomotive broadening in the harmonics when choosing the optimal delay. In this case, harmonics form a nice comb in the gamma region, and this result stands for different $a_0$ and $\tau$. Such effect may occur due to the fact that we are limiting harmonics emission to a quite narrow region around the peak of the pulse which means that 1) the intensity gradients are smaller, 2) the harmonics generation efficiency is higher (polarization is close to linear). Both of these lead to smaller ponderomotive broadening.


\begin{figure}

\centering
\includegraphics[width=0.85\linewidth]{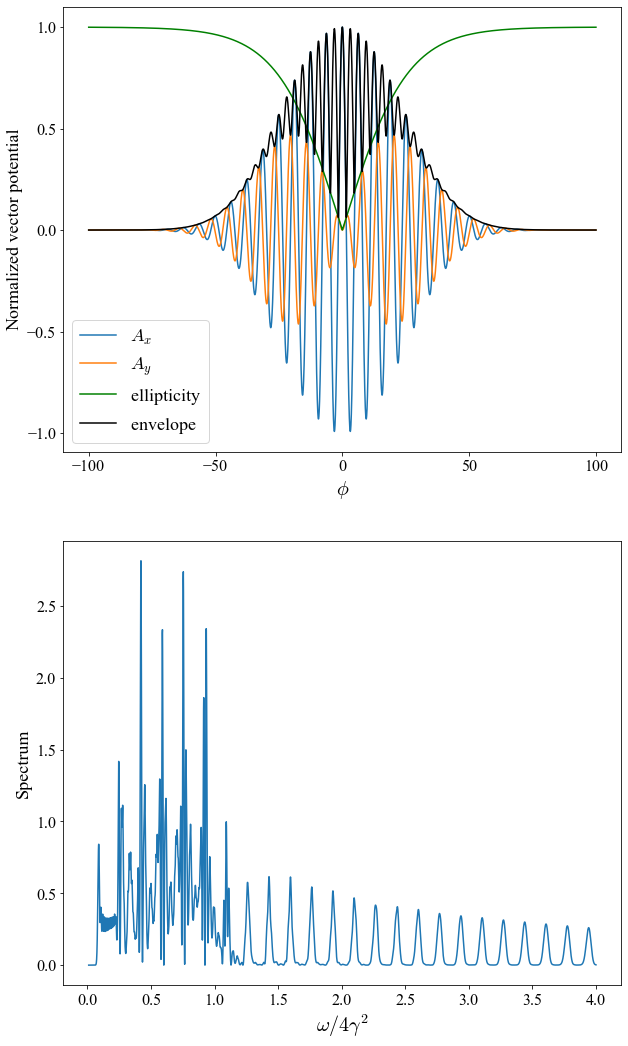}  

\caption{(Top) Vector potential and ellipticity of a polarization gated pulse for the optimal delay $\delta=\tau$. (Bottom) Corresponding backscattered spectrum. Vector potential and intensity are normalized. Laser pulse parameters: $a_0=3, \tau=8\pi$.}
\label{fig:A_and_back_spect_for_opt_delay}
\end{figure}

To investigate whether such comb pattern remains in the angular distribution, we calculated the spectrum dependence on the solid angle $\Omega$ and integrated it over the polar angle.
We observed that for the optimal delay parameter the comb could be still seen, although the more one goes away from the axis, the more broad and messy the harmonics comb will be. For delays that are not close to optimal, results seem to repeat the on-axis case: no distinct pattern in the harmonics spectrum was noticed. 

\begin{figure}
\centering
\includegraphics[width=.8\linewidth]{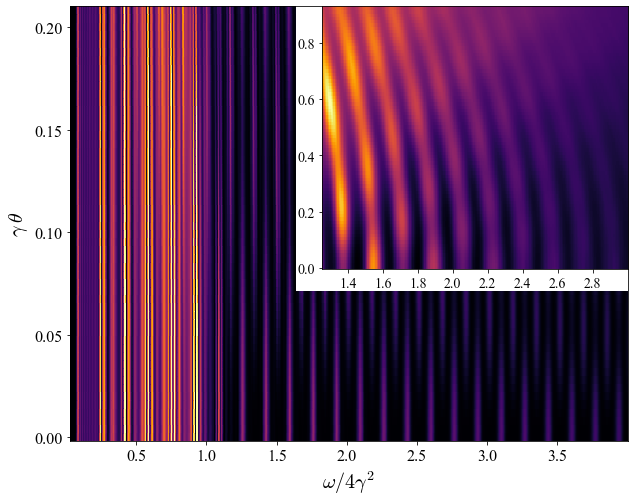}  
\caption{Radiation angular spectrum in the Lab frame for the optimal delay parameter $\delta=\tau$, $\gamma=529$. (Main) From one electron, $a_0=3, \tau=8\pi$. (Corner) From a realistic electron beam, $a_0=2, \tau=30\pi$.}
\label{fig:2D_lab}
\end{figure}

From the experimental point of view, it is especially interesting to discuss obtained results in the laboratory frame and whether a gamma comb could be detected. As it is well-known, Lorentz transforming back does not change the on-axis spectrum qualitatively (only the frequency is upshifted by $4\gamma^2$), therefore, the on-axis gamma comb remains. Figure \ref{fig:2D_lab} shows that the angular spectrum is squeezed into $1/\gamma$ cone but the pattern is still visible. 
In the close vicinity to the axis, one can obtain the harmonics comb for $\omega>4\gamma^2$ and directly measure it in experiments.

We would also like to repeat that the polarization gating technique is not aimed to avoid the ponderomotive broadening around the fundamental Compton line which could be noticed from both figures. In the electron's rest frame the main line is broadened up to $\omega = 1$ and interferes with harmonics falling into this interval, that is why only from $\omega \sim 1$ the effect shows itself.

\begin{figure}
\centering
\includegraphics[width=.8\linewidth]{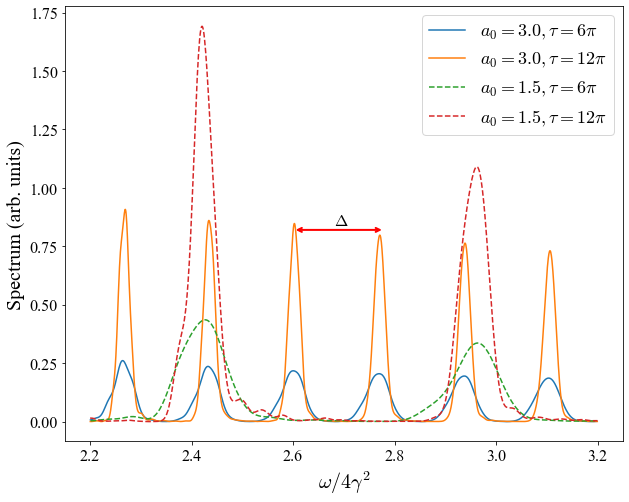}  
\caption{Normalized backscattered harmonics spectrum of an optimal polarization gated pulse for different pulse lengths: $6\pi, 12\pi$; and strengths $a_0=1.5,3$. Red arrow shows the distance between the 31st and 33rd harmonic.}
\label{fig:harmonics_for_different_lengths}
\end{figure}

Such comb has two characteristic properties: the distance between two adjacent peaks and the width of each harmonic, which could be controlled by the strength and length of the incident pulse ($a_0$ and $\tau$). Figure \ref{fig:harmonics_for_different_lengths} shows the normalized backscattered spectrum in the gamma region of optimal PGPs for different $a_0=1.5,\:3$ and $\tau=6\pi,\: 12\pi$. The distance between two peaks could be estimated from Eqs. (1)-(2) as $\Delta = 2 / (1 + 2 a_0^2 e^{-1/2})$ (the exponential factor is due to the gaussian temporal envelope) and is governed solely by $a_0$. Therefore, more intense laser pulses form more frequent gamma combs. As for the harmonic width, for longer laser pulses the comb is narrower (see Figure \ref{fig:harmonics_for_different_lengths}) as well as for more intense ones.

\begin{figure}
\centering
\includegraphics[width=.8\linewidth]{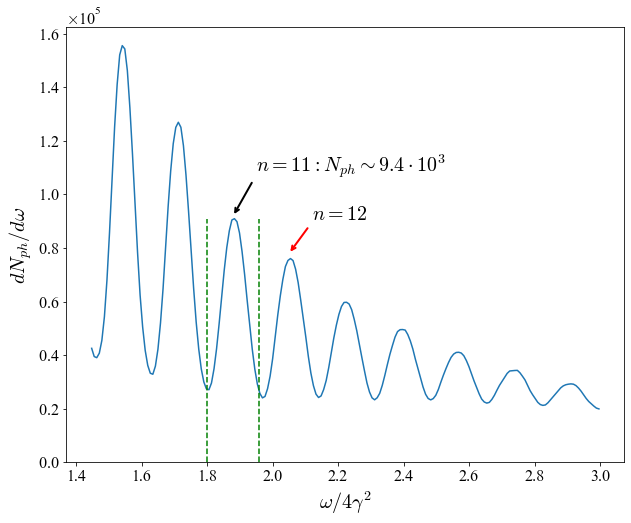}  
\caption{The differential number of photons in the gamma region scattered from a realistic electron beam ($10^8$ electrons represented by $4800$ macro particles, $\gamma=529$). Laser pulse parameters: $a_0=2, \tau=30\pi$.}
\label{fig:beam_n_ph_11}
\end{figure}

It is quite interesting to calculate how many photons are emitted into one particular harmonic. The exact photon number in the desired bandwidth $\frac{d^2 N_{ph}}{d\omega d\Omega} = \alpha \frac{1}{\omega} \frac{d^2 I}{d\omega d\Omega}$, where $\alpha \approx 1/137$ is the fine structure constant, could be estimated by integrating the photon spectrum over the proper collimation angle. The harmonics frequency is known from Eq. (2), and its approximate width could be numerically estimated. For instance, if we scatter a plane wave ($a_0=3, \tau=8\pi$) on a single electron ($\gamma=529$), integrate the differential number of photons over the collimation angle $\theta_{col} = 0.2/\gamma$, then the number of photons in the 23rd harmonic in the bandwidth $\frac{\Delta \omega}{\omega} \approx 0.04$ is around $1.9 \cdot 10^{-5}$.
In order to estimate whether a gamma comb could be observed in the real-life experimental setup, we simulated the interaction of a non-ideal electron beam with a polarization gated pulse using the code VDSR \cite{chen2013modeling}. The electron beam had realistic parameters: central gamma factor $\gamma=529$, normalized emittance $\epsilon_n \approx 0.14$ mm mrad, transverse radius $\sigma_{r} \approx 1.4 \: \mu$m, angular divergence $\sigma_{\theta} \approx 0.19 $ mrad, energy divergence $\approx 1\%$. The laser pulse was simulated in the paraxial approximation with $\lambda_0 = 0.8\: \mu$m and spotsize $w_0 = 24\: \mu$m. Figure \ref{fig:beam_n_ph_11} shows the differential number of photons in the gamma region integrated over the collimation angle $\theta_{col}=0.2/\gamma$ scattered from an electron beam ($10^8$ electrons represented by $4800$ macro particles). For instance, there are $\sim10^4$ photons in the 11th harmonic. Due to the relatively large collimation angle and electron beam non-ideal effects, one can observe even, off-axis harmonics between the gamma comb peaks. We can see that due to the broadening caused by the beam's angular and energy divergence \cite{rykovanov2014quasi}, harmonics start to overlap, but nevertheless, the nearest harmonics are still distinctly seen while the highest harmonics are more blurred. The reason is that the highest harmonics are less intensive, therefore not so noticeable against the background. This particular simulation shows that the gamma comb could be observed using a compact setup employing a laser system driving both the LPA and backscattering. One should mention that the photon numbers in a single narrow bandwidth line in the polarization gating method are comparable to those in conventional Compton backscattering facilities~\cite{weller2009research}. This makes our proposed scheme feasible for future photo-nuclear physics experiments.

In strong-field QED, nonlinear Compton scattering is described as a first-order process in the Furry picture using known solutions of the Dirac equation for the dressed electrons in the plane wave - Volkov spinor wave functions $\Psi_{p,\sigma}$ with asymptotic four-momentum $p$ and spin-polarization $\sigma$ \cite{kharin2018higher}. We also checked that in numerical simulations based on QED description the gamma comb is present.



Overall, we proposed a polarization gating technique - an experimentally feasible and simple method for avoiding the ponderomotive broadening (caused by temporal envelope) in the harmonics spectrum. We showed that for the optimal delay between circular pulses (which equals pulse length) one can observe a narrow bandwidth comb in the gamma region for the backscattered spectrum as well as for angular spectrum. Such an effect may arise due to the fact that we limit harmonics emission to the region around the pulse's peak, where the harmonics emission efficiency is the highest and intensity gradients are the smallest, which significantly reduces ponderomotive broadening. One can change the laser pulse intensity and length to control how frequent and narrow the gamma comb will be. By choosing a proper collimation angle one may estimate the number of photons in a particular harmonic. We simulated the interaction of the polarization gated pulse with a realistic electron beam and showed that the gamma comb could still be observed in the real-life experimental setup. For the proof-of-principle experiments, it would make sense to perform a fully optical experiment using an LPA and a scattering laser. However, for the usage of the proposed gamma comb as a bright source, it seems to be better to use conventional accelerators due to their superior longitudinal and transverse emittance. Our proposed scheme might be useful in photo-nuclear experiments as well as nonlinear QED experiments planned at DESY (LUXE~\cite{abramowicz2021conceptual, kampfer2021impact}) and SLAC (experiment E-320).

The authors acknowledge the usage of Skoltech CDISE supercomputer ``Zhores''~\cite{zacharov2019zhores} for obtaining the numerical results presented in this paper. S.R. would like to thank V.G. Nedorezov for fruitful discussions.



\providecommand{\noopsort}[1]{}\providecommand{\singleletter}[1]{#1}%

\end{document}


\maketitle

\section{PGP with carrier-envelope effects}
\begin{figure}[h!]
    \centering
    \includegraphics[width=0.5\textwidth]{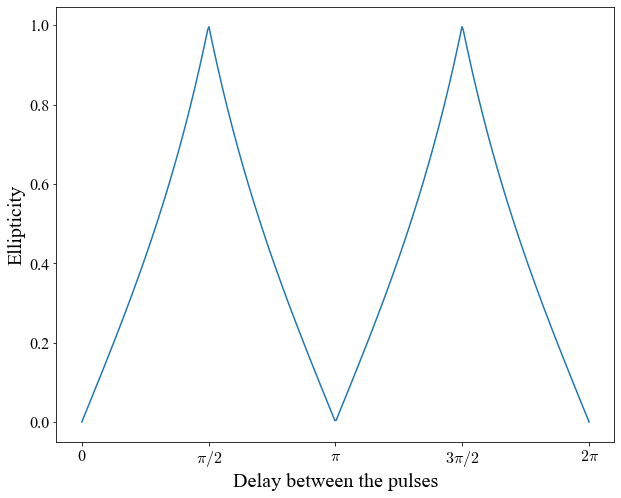}
    \caption{Ellipticity at the centre of a PGP as a function of the delay between two circular pulses modulo $2\pi$. The carrier-envelope phase effects are taken into account. Ellipticity $=1$ corresponds to circular polarization, while ellipticity $=0$ - to linear. One may notice that only for delays that are multiples of $\pi$ the ellipticity at the centre of the pulse ($\phi=0$) is linear.}
    \label{fig_appex:ell_vs_delay}
\end{figure}

\noindent The 3-vector potential $\mathbf{A}=(A_x,A_y,0)$ of a PGP with the carrier-envelope phase is given by
\begin{align*}
    &A_x(\phi) = a_0 \left[\:g(\phi + \delta/2) \cos(\phi + \delta/2) + g(\phi - \delta/2) \cos(\phi - \delta/2)\:\right], \\
    &A_y(\phi) = a_0 \left[\:g(\phi + \delta/2) \sin(\phi + \delta/2) - g(\phi - \delta/2) \sin(\phi - \delta/2)\:\right],
\end{align*}
which at $\phi=0$ for a symmetric envelope $g(\cdot)$ reduces to
\begin{align*}
    &A_x(0) = 2 \:a_0\: g(\delta/2) \cos(\delta/2), \\
    &A_y(0) = 2 \:a_0\: g(\delta/2) \sin(\delta/2).
\end{align*}
Therefore, if we take into account the carrier-envelope phase, the delay should be a multiple of $\pi$ in order to have linear polarization at the centre (see Figure \ref{fig_appex:ell_vs_delay}). If we do not take into account the carrier-envelope phase effects, then for any symmetric pulse the polarization at the centre is linear.

\section{Classical simulations with radiation reaction and \\QED simulations}

In the main text of the paper, we considered the classical motion of the particle in a given background field not taking into account the back-reaction of the scattered radiation. Here we provide the results of numerical simulation taking radiation
reaction into account.

We approach the problem in two ways. Classically radiation reaction can be included in consideration using the Landau-Lifschitz equation. The equation admits an analytical solution for the case of a plane wave field \cite{di2008exact}. The solution can be plugged into master-integral for calculating far-field emission spectrum for a given particle trajectory.

Quantum electrodynamics, on the other hand, also provides a way to incorporate radiation reaction effects. They arise as a consequence of momentum conservation when taking emitted photon momentum into account. In order to obtain numerical results, we considered first-order transition amplitudes between Volkov states. Volkov states are exact solutions of the Dirac equation in the presence of plane wave classical background field. The presence of quantized field adds perturbation to Lagrangian and makes the transitions between Volkov states possible with photon emission. Corresponding probabilities were summed over emitted photon polarizations and averaged over initial electron spin as in \cite{kharin2018higher}.

\begin{figure}
    \begin{subfigure}{0.45\textwidth}
    \centering
    \includegraphics[width=\textwidth]{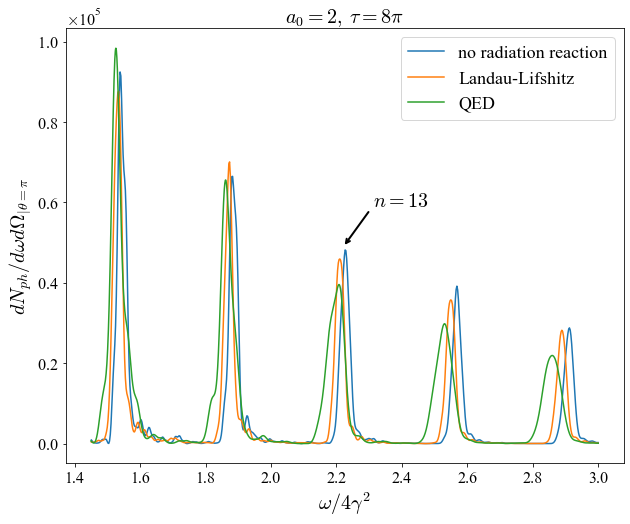}
    \end{subfigure}
    \hspace{16pt}
    \begin{subfigure}{0.45\textwidth}
    \centering
    \includegraphics[width=\textwidth]{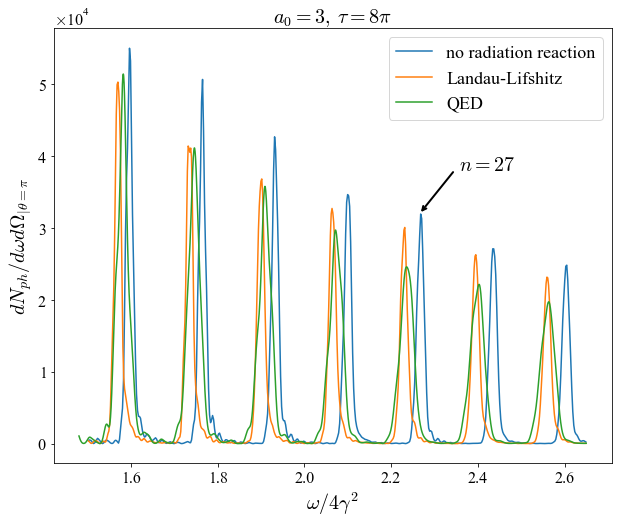}
    \end{subfigure}
    \caption{Differential number of photons per unit frequency emitted on axis from one electron off an optimal PGP with (Left) $a_0=2\: \tau=8\pi$, (Right) $a_0=3\: \tau=8\pi$. The calculations were done via (blue) calculation of spectrum integral in Eq. (5), (yellow) integration of Landau-Lifshitz equation and (green) QED simulations.}
    \label{fig_appex:n_ph_clas_LL_qed}
\end{figure}

In both cases one can see the redshift of harmonics in comparison with the case described initially (see Fig. \ref{fig_appex:n_ph_clas_LL_qed}). Classically, that is the result of ``radiation friction'', and in quantum that is due to recoil. However, with the parameters described in the paper, the difference is small and does not affect the key idea.

\section{Optimal PGP and delay variation}

To clearly demonstrate the presence of the gamma comb for delays that are close to optimal, we neglected the carrier-envelope phase effects and modeled several PGPs with delays greater and smaller than optimal as well as the optimal one (see Figure \ref{fig_appex:ell_vs_delay}).

Figure S.4 shows backscattered spectra from the optimal PGP and regular gaussian pulse. One may notice that for the optimal PGP the harmonics are narrow, while the main emission line remains broad.

Figure S.5 shows the differential number of photons in the harmonics region scattered from a single electron.

\begin{figure}[h!]

\centering
\begin{subfigure}{\textwidth}
\includegraphics[width=0.9\linewidth]{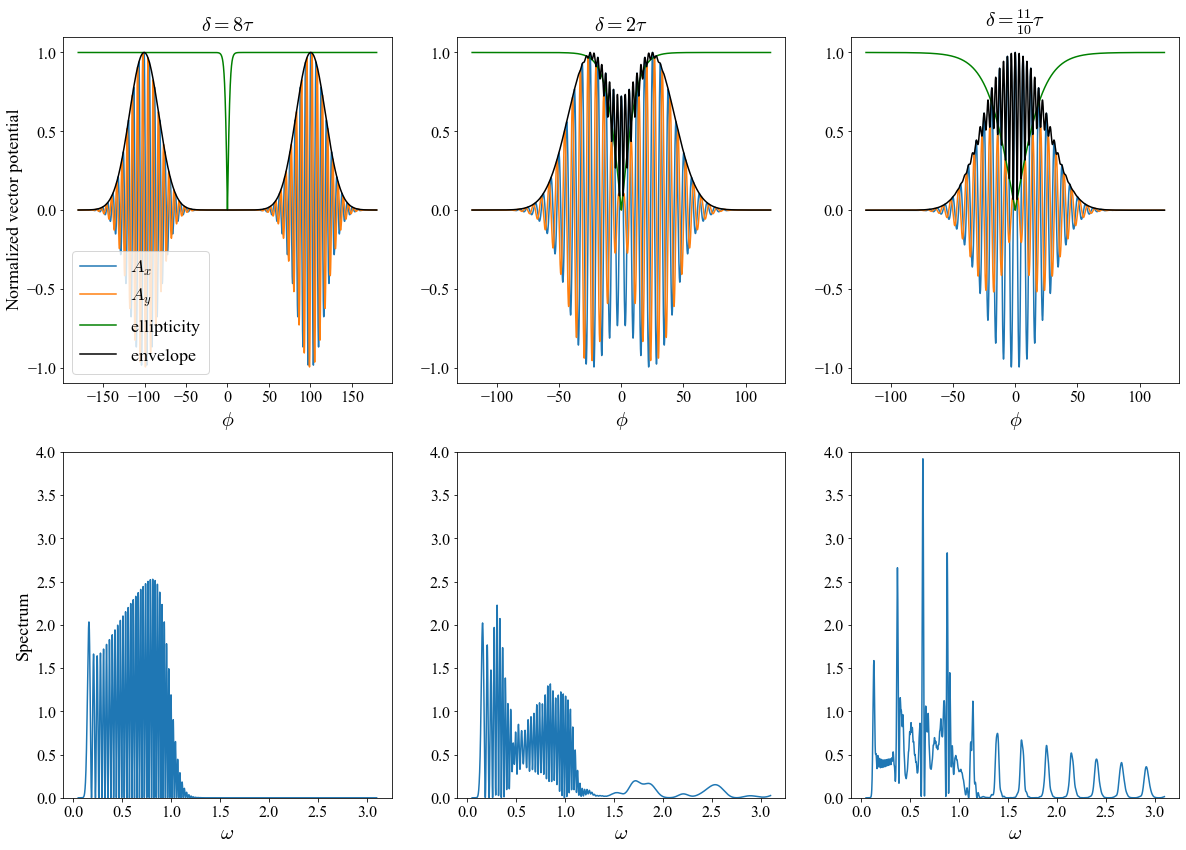} 
\end{subfigure}

\begin{subfigure}{\textwidth}
\includegraphics[width=0.9\linewidth]{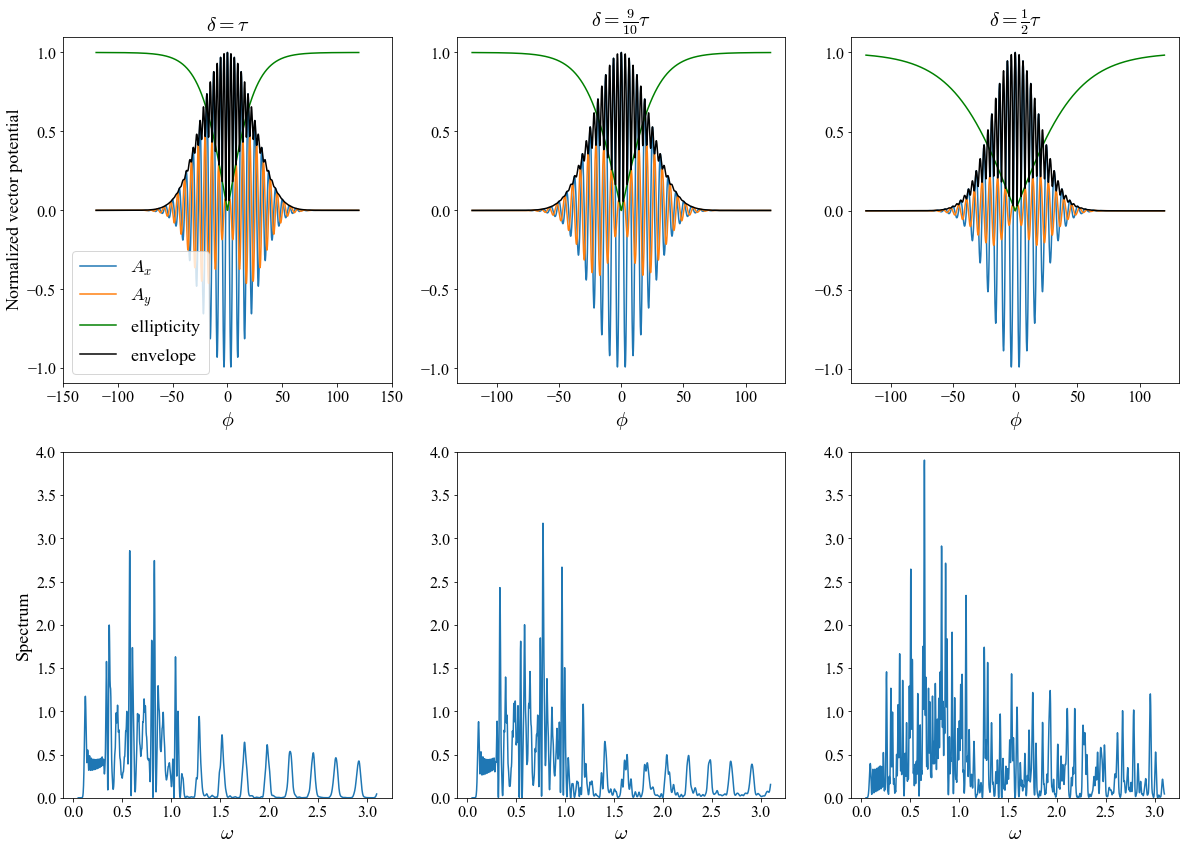} 
\end{subfigure}

\caption{(Top) Vector potential and ellipticity of polarization gated pulses with different delays. (Bottom) Corresponding backscattered spectrum. Presented delays: $\delta=8\tau, 2\tau, \frac{11}{10}\tau, \tau, \frac{9}{10}\tau,  \frac{1}{2}\tau$. Laser pulse parameters: $a_0=3, \tau=8\pi$. The vector potential is normalized. To obtain the spectrum in ergs one needs to additionally multiply values on the figure by the normalization coefficient $e^2 \omega_L$. Here we neglect carrier-envelope phase effects, therefore, non-multiple of $\pi$ delays still result in linear polarization at the center of the pulse.}
\label{fig_appex:A_different_delta}
\end{figure}

\begin{figure}[h!]
\begin{subfigure}{.45\textwidth}
\centering
\includegraphics[width=\textwidth]{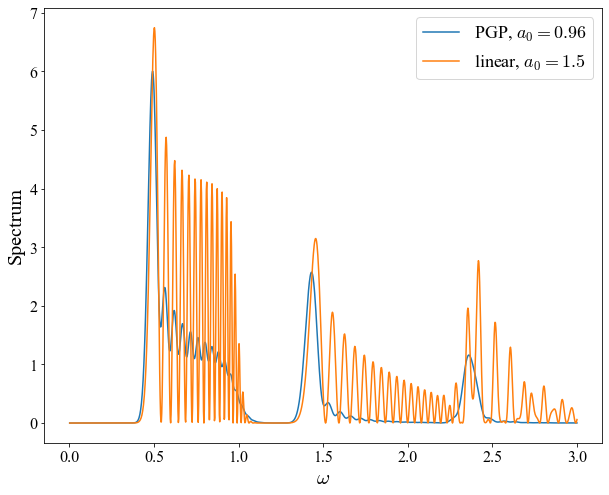}  
\caption{Figure S.4: Backscattered spectrum from an optimal PGP and linear pulse of the same energy. PGP parameters: $a_0=0.96, \tau=10\pi$. Linear pulse parameters: $a_0=1.5, \tau=20\pi$.}
\label{fig_appex:pol_gate_vs_linear}
\end{subfigure}
\hspace{16pt}
\begin{subfigure}{.45\textwidth}
\centering
\includegraphics[width=\linewidth]{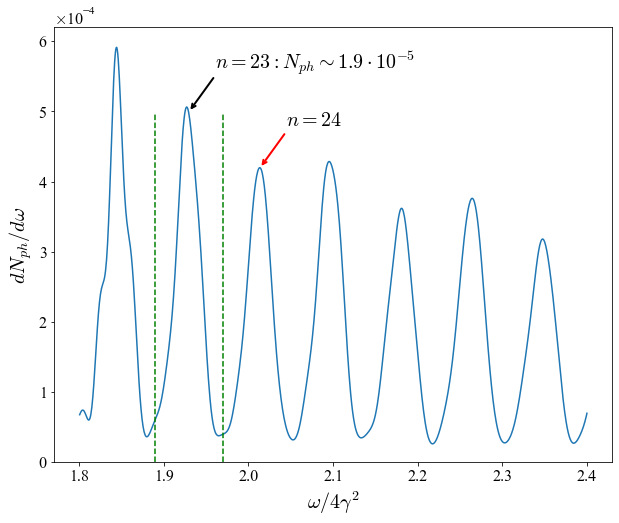}  
\caption{Figure S.5: Differential number of photons in the gamma region scattered from one electron ($\gamma=529$). Laser pulse parameters: $a_0=3, \tau=8\pi$.}
\label{fig_appex:n_ph_1}
\end{subfigure}
\end{figure}

\clearpage





\section{Analytics on polarization gating technique}

\subsection{Goal}
The purpose of the present document is to gain justified intuition behind the
proposed polarization gating technique. Namely, explain the origin of optimal
delay in order to provide guidance for estimates on optimal parameters and
stability.

\subsection{Preliminaries}
We start with the Lienard--Wiechert potentials and the spectrum of radiation.
Photon density of the emitted radiation (with wave vector $k$, polarization
vector $\varepsilon$) is given by
\begin{equation}
  \frac{d^2 N_\varepsilon}{d\omega\,d\Omega} = \frac{\alpha\omega}{4\pi^2}
  \left| \int_{-\infty}^{\infty} e^{ik\cdot x} d(\varepsilon\cdot x)\right|^2\,.
  \label{eq:master}
\end{equation}

Here $\alpha$ is a fine-structure constant, $x^\mu(s)$ is the particle
trajectory, $\omega\equiv k^0$ is the scattered photon frequency. $\omega$ and
$x$ are taken in reciprocal dimensionality, solid angle element $d\Omega$ is
dimensionless.

Suppose the wave is propagating in $x^3$ direction. By the symmetries with
respect to translations $x^1\mapsto x^1+a^1$, $x^2\mapsto x^2+a^2$ and $x^+
\equiv x^0+x^3 \mapsto x^+ + a^+$, the corresponding conjugate momenta are
conserved. Choose the gauge where potential of the incoming wave has the
components $A^1(x^-)$ and $A^2(x^-)$ only. Suppose also that particle was
initially at rest (one can always achieve that by appropriate choice of the
reference frame). Four-velocity of the particle obeys
\begin{align}
  u^{1,2} &= A^{1,2}\,,\\
  u^{0} - u^{3} &= 1\,,\\
  u^{0} + u^{3} &= 1+ \frac{\mathbf{A}^2}{2}\,.
  \label{eq:fourvel}
\end{align}
Here we work in natural units $c=\hbar=1$, we also incorporated particle charge
and mass in the vector-potential. That is, we work with the units of $a_0$ now.

Consider for simplicity the case of back-scattering. The corresponding photon
densities take the form ($j=1$ and $j=2$ stand for the $x$- and $y$-
polarizations)
\begin{equation}
  \frac{d^2 N_j}{d\omega\,d\Omega} = \frac{\alpha\omega}{4\pi^2}
  \left| \int_{-\infty}^{\infty} A^j(x^-) \exp\left\{i\omega\int_{-\infty}^{x^-}
  \left[1+\frac{\mathbf{A}(s)^2}{2}\right]ds\right\} d x^-\right|^2\,.
  \label{eq:master2}
\end{equation}

\subsection{Expressions for polarization gating}
We focus on the pulses with slowly varying amplitude. We also assume the delay
to be contain an integer number of wavelengths. The absolute phase of the pulse
is irrelevant to our discussion. Therefore, by the appropriate choice of
coordinates, the vector potential of incoming wave can be written as
\begin{align}
  A^1 &= (g_+ + g_-) \cos x^-\,,\\
  A^2 &= (g_+ - g_-) \sin x^-\,,\\
  g_{\pm} &\equiv a\left(x^-\pm \frac{\delta}{2}\right)\,.
  \label{eq:pg_wave}
\end{align}
Here $a(x^-)$ is an envelope of a single pulse, $\delta$ is the delay. Applying
slowly varying amplitude approximation to the expression (\eqref{eq:master2}),
and using Jacobi-Anger decomposition yields
\begin{equation}
  \frac{d^2 N_j}{d\omega\,d\Omega} \approx
  \frac{\alpha\omega}{4\pi^2}
  \left|\sum_{m=-\infty}^{\infty} 
  \int_{-\infty}^{\infty} A^j(x^-) 
  J_{m}\left(\frac{\omega g_+ g_-}{2}
  \right)
  \exp\left\{i(\omega+2m) x^-+\frac{i\omega}{2}\int_{-\infty}^{x^-}
  g_+(s)^2+g_-(s)^2 ds\right\} d x^-\right|^2\,.
  \label{eq:decomp1}
\end{equation}
Denote
\begin{equation}
  \psi_{2m+1}(x^-)\equiv [\omega - (2m+1)]x^- +\frac{\omega}{2}
  \int_{-\infty}^{x^-}
  [g_+(s)^2+g_-(s)^2] ds\,,
  \label{eq:expphase}
\end{equation}
And
\begin{align}
  P_{m}^1(x^-)&\equiv
  (-1)^m \left[
    J_{m+1}\left(\frac{\omega g_+ g_-}{2}\right) - 
    J_{m}\left(\frac{\omega g_+ g_-}{2}\right)
  \right] (g_+ + g_-)\,\\
  P_{m}^2(x^-)&\equiv
  (-1)^m \left[
    J_{m+1}\left(\frac{\omega g_+ g_-}{2}\right) + 
    J_{m}\left(\frac{\omega g_+ g_-}{2}\right)
  \right] (g_+ - g_-)\,.
  \label{eq:expfactor}
\end{align}
Note $\psi$ being the rapidly changing phases (its stationry points define the
harmonics), and $P$ being the slowly-varying prefactors.
\begin{equation}
  \frac{d^2 N_j}{d\omega\,d\Omega} \approx \frac{\alpha\omega}{16\pi^2}
  \left|\sum_{m=-\infty}^{\infty} 
  \int_{-\infty}^{\infty} P_m^{j}(x^-)
  e^{i\psi_{2m+1}(x^-)}
  d x^-\right|^2\,.
  \label{eq:decomp2}
\end{equation}
Every term corresponds to the harmonic of the order $2m+1$ observed in on-axis
scattering. Evaluating the stationary phase in the exponential, one can see that
the ``instantaneous'' emitted frequency $\omega^*$ obeys 
\begin{equation}
  \omega^*(x^-) = \frac{2m+1}{1 + \frac{g_+^2+g_-^2}{2}}\,.
  \label{eq:inst_freq}
\end{equation}

\subsection{Interpretation}
Consider higher harmonics ($m>0$). We see that there are two ``effects''
addressing the shape of every harmonic. One is strobing originating from the
prefactors $P$, and another one is that different parts of the pulse contribute
to different frequency. Passing to Stokes parameters of the incident wave
provides nice intuition behind that. The strobing is due to polarization
effects, whereas the frequency shift remains dependent solely on radiation
intensity.

\subsection{Gaussian pulses}
Note the product $g_+ g_-$ in the argument of Bessel function. If we assume both
pulses to be Gaussian, the ``width'' of this product is always the same,
independent on the inter-pulse delay. Therefore, one can achieve polynomial
suppression of the harmonics out of the overlap region. On the other hand, the
character of the frequency shift highly depends on the delay. In order for
overlap region to contribute to the same frequency, one may wish $g_+^2 + g_-^2$
to be constant in this region. Taking both the pulses Gaussian, this would mean
$d^2(g_+^2+g_-^2)/d\,(x^-)^2=0$ at $x^-=0$. Take $g(x)=a_0e^{-x^2/\tau^2}$. This
immediately leads to $\delta = \tau$.

\printbibliography